# Quantum mechanics with orthogonal polynomials

A. D. Alhaidari

*Saudi Center for Theoretical Physics, P.O. Box 32741, Jeddah 21438, Saudi Arabia*

**Abstract**: We present a formulation of quantum mechanics based on orthogonal polynomials. The wavefunction is expanded over a complete set of square integrable basis in configuration space where the expansion coefficients are orthogonal polynomials in the energy. Information about the corresponding physical systems (both structural and dynamical) are derived from the properties of these polynomials. We demonstrate that an advantage of this formulation is that the class of *exactly solvable* non-relativistic quantum mechanical problems becomes larger than in the conventional formulation (see, for example, Table 1 in the text).

## 1. Introduction

In physics, we are accustomed to writing vector quantities (e.g., force, velocity, electric field, etc.) in terms of their components in some conveniently chosen vector space. For example, the force $\vec{F}$ is written in three dimensional space with Cartesian coordinates as $\vec{F} = f_x\hat{x} + f_y\hat{y} + f_z\hat{z}$, where $\{f_x, f_y, f_z\}$ are the components of the force along the unit vectors $\{\hat{x}, \hat{y}, \hat{z}\}$. These components contain all physical information about the quantity whereas the unit vectors (basis) are dummy, but must form a complete set to allow for a faithful physical representation. In fact, we can as well write the force in another coordinates, say the spherical coordinates with basis $\{\hat{r}, \hat{\theta}, \hat{\varphi}\}$, as $\vec{F} = f_r\hat{r} + f_\theta\hat{\theta} + f_\varphi\hat{\varphi}$, where $\{f_r, f_\theta, f_\varphi\}$ are the new components that contain the same physical information. And so on, where in general we can write $\vec{F} = \sum_n f_n\hat{x}_n$. The basis (unit vectors) $\{\hat{x}_n\}$ is chosen conveniently depending on the symmetry of the problem (e.g., rectangular, spherical, cylindrical, elliptical, etc.). In fact, it does not even have to be orthogonal.

In quantum mechanics we can also think of the wavefunction $|\psi(x)\rangle$ as a local vector and write it in terms of its components $\{f_n\}$ along some local unit vectors (basis) $\{|\phi_n(x)\rangle\}$ as $|\psi(x)\rangle = \sum_n f_n|\phi_n(x)\rangle$. All physical information about the system are contained in the components (expansion coefficients) $\{f_n\}$. On the other hand, the basis set $\{\phi_n(x)\}$ is dummy but, in analogy to the unit vectors in the above example, must be normalizable (square integrable) and complete. If the physical system is also associated with a set of real parameters $\{\mu\}$, then the components of the wavefunction at an energy $E$ could be written as parameterized functions of the energy, $\left\{f_n^\mu(E)\right\}$, and the state of the system becomes

$$\psi_E^\mu(x)=\sum_n f_n^\mu(E)\,\phi_n(x)\,. \tag{1}$$

We have shown elsewhere [1] that if we write $f_n^\mu(E)=f_0^\mu(E)P_n^\mu(\varepsilon)$, where $\varepsilon$ is some proper function of $E$ and $\{\mu\}$, then completeness of the basis and energy normalization of the density of state make $\left\{P_n^\mu(\varepsilon)\right\}$ a complete set of orthogonal polynomials with an associated positive weight function $\rho^\mu(\varepsilon)=[f_0^\mu(E)]^2$. That is,

$$\int\rho^\mu(\varepsilon)P_n^\mu(\varepsilon)P_m^\mu(\varepsilon)\,d\zeta(\varepsilon)=\delta_{n,m}\,, \tag{2}$$

where $d\zeta(\varepsilon)$ is an appropriate energy integration measure. Therefore, we can rewrite the wavefunction expansion (1) as follows

$$\psi_E^\mu(x)=\sqrt{\rho^\mu(\varepsilon)}\sum_n P_n^\mu(\varepsilon)\,\phi_n(x)\,. \tag{3}$$

Scattering states dictate that the physically relevant polynomials are only those with the following asymptotic ($n\to\infty$) behavior [1-3]

$$P_n^\mu(\varepsilon)\approx n^{-\tau}A^\mu(\varepsilon)\cos[n^\xi\,\theta(\varepsilon)+\delta^\mu(\varepsilon)]\,, \tag{4}$$

where $\tau$ and $\xi$ are real positive constants. $A^\mu(\varepsilon)$ is the scattering amplitude and $\delta^\mu(\varepsilon)$ is the phase shift. Bound states, if they exist, occur at a set (infinite or finite) of energies that make the scattering amplitude vanish. That is, the $k^{\text{th}}$ bound state occurs at an energy $E_k=E(\varepsilon_k)$ such that $A^\mu(\varepsilon_k)=0$ and the corresponding bound state wavefunction is written as

$$\psi_k^\mu(x)=\sqrt{\omega^\mu(\varepsilon_k)}\sum_n Q_n^\mu(\varepsilon_k)\,\phi_n(x)\,, \tag{5}$$

where $\left\{Q_n^\mu(\varepsilon_k)\right\}$ are the discrete version of the polynomials $\left\{P_n^\mu(\varepsilon)\right\}$ and $\omega^\mu(\varepsilon_k)$ is the associated discrete weight function. That is, $\sum_k\omega^\mu(\varepsilon_k)Q_n^\mu(\varepsilon_k)Q_m^\mu(\varepsilon_k)=\delta_{n,m}$. Sometimes, the quantum system consists of continuous as well as discrete energy spectra simultaneously. In that case, the wavefunction is written as follows

$$\psi_k^\mu(x,E)=\sqrt{\rho^\mu(\varepsilon)}\sum_n P_n^\mu(\varepsilon)\,\phi_n(x)+\sqrt{\omega^\mu(\varepsilon_k)}\sum_n P_n^\mu(\varepsilon_k)\,\phi_n(x)\,. \tag{6}$$

In such cases, the appropriate polynomial orthogonality becomes

$$\int\rho^\mu(\varepsilon)P_n^\mu(\varepsilon)P_m^\mu(\varepsilon)\,d\zeta(\varepsilon)+\sum_k\omega^\mu(\varepsilon_k)P_n^\mu(\varepsilon_k)P_m^\mu(\varepsilon_k)=\delta_{n,m}\,, \tag{7}$$

In the conventional (textbook) formulation of quantum mechanics, the potential function plays a central role in providing physical information about the system. Hence, in the present formulation, the set of orthogonal polynomials replaces the potential function in this role. In fact, it is more than just that. As we shall see below, the orthogonal polynomials also carry kinematic information (e.g., the angular momentum) whereas the potential function does not.

Since the orthogonal polynomials (both continuous and discrete) satisfy three-term recursion relations, then the basis set $\{\phi_n(x)\}$ is required to produce a tridiagonal matrix representation for the corresponding wave operator. As such, the matrix wave equation becomes equivalent, and amenable to the said recursion relation. In this paper, we leave out technical details but include the most relevant information concerning orthogonal polynomials in the Appendices. Interested readers may consult cited references for the derivation of applicable results, especially Ref. [1] and Ref. [4].

In the following three sections, we present examples of known orthogonal polynomials of the hypergeometric type and derive the properties of the corresponding physical systems. Finally, we conclude in section 5 by making relevant comments and discussing related issues. Throughout the paper, we adopt the atomic units, $\hbar = M = 1$.

# 2. The Meixner-Pollaczek polynomial class of problems

In this section, we consider the class of quantum systems whose continuum scattering states are described by the wavefunction (3) where the expansion coefficients are the Meixner-Pollaczek polynomials $P_n^{\mu}(z,\theta)$ shown in Appendix A by Eq. (A1).

### 2.1 The Coulomb Problem

We start by choosing the physical parameters in the polynomial as: $\mu = \ell + 1$, $\cos\theta = \frac{2E-(\lambda/2)^2}{2E+(\lambda/2)^2}$ and $z = Z/\sqrt{2E}$, where $\ell$ is a non-negative integer, $Z$ is a real number and $\lambda$ is a positive length scale parameter. Depending on the range of values of the physical parameters, this system can have a continuous or discrete energy spectrum. For example, if $E$ is negative then $z$ becomes pure imaginary and $|\cos\theta| > 1$. As explained in Appendix A, this is equivalent to the replacement of $z \to \mathrm{i}z$ and $\theta \to \mathrm{i}\theta$, which turns the Meixner-Pollaczek polynomials into its discrete version, the Meixner polynomials. Substituting these parameters in Eq. (A7) gives the following scattering phase shift

$$\delta(E) = \arg\Gamma\left(\ell + 1 + \mathrm{i}\,Z/\sqrt{2E}\right). \tag{8}$$

Whereas, substitution in Eq. (A8) gives the following energy spectrum formula

$$E_k = -\frac{1}{2}\frac{Z^2}{\left(k+\ell+1\right)^2}. \tag{9}$$

These results are identical to those of the well-known Coulomb interaction in three dimensions, $V(r) = Z/r$, where $r$ is the radial coordinate, $Z$ is the electric charge and $\ell$ is the angular momentum quantum number. The discrete polynomial in the bound states wavefunction expansion (5) is the Meixner polynomial shown in Appendix A by Eq. (A9). The requirement on the basis to support a tridiagonal matrix representation of the Schrödinger wave operator, $-\frac{1}{2}\frac{d^2}{dr^2} + \frac{\ell(\ell+1)}{2r^2} + \frac{Z}{r} - E$, gives (see section II.A.2 in Ref. [4])

$$\phi_n(r) = \sqrt{\tfrac{\Gamma(n+1)}{\Gamma(n+2\ell+2)}}\,(\lambda r)^{\ell+1} e^{-\lambda r/2} L_n^{2\ell+1}(\lambda r), \tag{10}$$

where $L_n^{\nu}(z)$ is the Laguerre polynomial. This is not the only exactly solvable problem in the Meixner-Pollaczek polynomial class. Next, we give two other examples; one with only an infinite bound states spectrum and another with a continuous as well as discrete finite spectrum (however, in this class it is exactly solvable only for the discrete bound states).

## 2.2 The oscillator problem

For the second example in this class, we take the polynomial parameters as $\mu=\frac{1}{2}\left(\ell+\frac{3}{2}\right)$ and $z=\mathrm{i}E/2\omega$, where $\omega$ is a real number. The choice of $z$ as pure imaginary mandates the replacement $\theta\to\mathrm{i}\theta$ so that reality is maintained for the polynomial (A1) and its recursion relation (A4). As seen in Appendix A, this leads only to bound states and the discrete Meixner polynomial. The infinite energy spectrum formula is obtained from Eq. (A8) as

$$E_k=\omega\left(2k+\ell+\tfrac{3}{2}\right). \tag{11}$$

This corresponds to the well-known energy spectrum of the isotropic oscillator $V(r)=\frac{1}{2}\omega^2r^2$ with oscillator frequency $\omega$. The corresponding eigenstates are written as in Eq. (5) in terms of the discrete Meixner polynomial. Moreover, the requirement that the basis yield a tridiagonal matrix representation for the wave operator, $-\frac{1}{2}\frac{d^2}{dr^2}+\frac{\ell(\ell+1)}{2r^2}+\frac{1}{2}\omega^2r^2-E$, gives (see section II.A. 1 in Ref. [4])

$$\phi_n(r)=\sqrt{\tfrac{\Gamma(n+1)}{\Gamma(n+\ell+3/2)}}\,(\lambda r)^{\ell+1}e^{-\lambda^2r^2/2}L_n^{\ell+1/2}(\lambda^2r^2), \tag{12}$$

where $\lambda$ is a length scale parameter such that $\lambda^2\le 4\omega$. Additionally, the parameter $\beta$ in the Meixner polynomial (A9) is obtained as $\beta=e^{-2\theta}$ where $\cosh\theta=\dfrac{\omega^2+(\lambda/2)^4}{\omega^2-(\lambda/2)^4}$.

## 2.3 The Morse problem

The final problem in the Meixner-Pollaczek polynomial class corresponds to the parameter assignments: $\mu=\frac{1}{2}+\sqrt{-\varepsilon}$ and $z=\mathrm{i}u_1/2\sqrt{u_0}$, where all parameters are real with $\varepsilon<0$, $u_0>0$ and $u_1<0$. Thus, it is required that $\theta\to\mathrm{i}\theta$ in Eqs. (A1) and (A4) turning the polynomial into one of its two discrete versions. Formula (A8) gives the energy spectrum as

$$\varepsilon_k=-\left(k+\tfrac{1}{2}+u_1/2\sqrt{u_0}\right)^2, \tag{13}$$

where $k=0,1,..,N$ and $N$ is the largest integer less than or equal to $-u_1/2\sqrt{u_0}-\frac{1}{2}$. If we introduce an inverse length parameter $\alpha$ and write $E=\frac{1}{2}\alpha^2\varepsilon$ and $V_i=\frac{1}{2}\alpha^2u_i$, then we can rewrite the spectrum formula (13) as follows

$$E_k=-\tfrac{1}{2}\alpha^2\left(k+\tfrac{1}{2}+V_1/\alpha\sqrt{2V_0}\right)^2, \tag{14}$$

This, in fact, is the energy spectrum formula of the one-dimensional Morse potential $V(x)=V_0e^{2\alpha x}+V_1e^{\alpha x}$ where $-\infty<x<+\infty$ [5]. The corresponding bound states are written as in Eq. (5) in terms of the discrete version of the Meixner-Pollaczek polynomial with a finite spectrum, which is the Krawtchouk polynomial not the Meixner polynomial. The orthonormal version of this polynomial is given in Appendix A by Eq. (A11). The requirement that the corresponding basis gives a tridiagonal matrix representation for the wave operator, $-\frac{1}{2}\frac{d^2}{dx^2}+V_0e^{2\alpha x}+V_1e^{\alpha x}-E$, results in (see section II.A.3 in Ref. [4])

$$\phi_n(x)=\sqrt{\tfrac{\Gamma(n+1)}{\Gamma(n+\nu+1)}}\,y^{\nu/2}e^{-y/2}L_n^{\nu}(y), \tag{15}$$

where $y(x)=e^{\alpha x}$ and $\nu=\frac{2}{\alpha}\sqrt{-2E}$ . Moreover, the parameter $\gamma=e^{-2\theta}$ in the Krawtchouk polynomial is obtained from $\cosh\theta=\frac{2V_0+(\alpha/2)^2}{2V_0-(\alpha/2)^2}$ with $V_0\geq\alpha^2/8$.

In this class, we were able to obtain full solutions for two problems, the Coulomb and the isotropic oscillator. The latter has only discrete bound states whereas the former has both discrete bound states as well as continuum scattering states. Additionally, we were able to obtain only partial solution to the 1D Morse oscillator. We could find only the discrete bound states solution but not the continuum scattering states. In the following section, we will remedy that.

# 3. The continuous dual Hahn polynomial class of problems

In this section, we consider the class of problems whose continuum scattering states are described by the wavefunction (3) where the expansion coefficients are the continuous dual Hahn polynomial $S_n^\mu(z^2;a,b)$ shown in Appendix B by Eq. (B1). Throughout this section, we restrict our investigation to the special case where the two polynomial parameters *a* and *b* are equal.

### 3.1 The Morse problem

We start by choosing the polynomial parameters as: $\mu=\rho+\frac{1}{2}$, $a=b=\frac{\nu+1}{2}$ and $z^2=\varepsilon$. If $\rho>-\frac{1}{2}$, then $\mu$ is positive and we obtain only a continuous spectrum corresponding to scattering states with the phase shift given in Appendix B by Eq. (B7) as

$$\delta^\mu(\varepsilon)=\arg\Gamma\left(2i\sqrt{\varepsilon}\right)-\arg\Gamma\left(\rho+\tfrac{1}{2}+i\sqrt{\varepsilon}\right)-2\arg\Gamma\left(\tfrac{\nu+1}{2}+i\sqrt{\varepsilon}\right). \tag{16}$$

If we introduce an inverse length parameter $\alpha$ and write $2E=\alpha^2\varepsilon$ and $2V=\alpha^2\rho$, then this scattering phase shift reads as follows

$$\delta^\mu(\varepsilon)=\arg\Gamma\left(\tfrac{2i}{\alpha}\sqrt{2E}\right)-\arg\Gamma\left(\tfrac{2V}{\alpha^2}+\tfrac{1}{2}+\tfrac{i}{\alpha}\sqrt{2E}\right)-2\arg\Gamma\left(\tfrac{\nu+1}{2}+\tfrac{i}{\alpha}\sqrt{2E}\right). \tag{17}$$

This is identical to the scattering phase shift associated with the 1D Morse potential, $V(x)=\frac{1}{2}\left(\frac{\alpha}{2}\right)^2e^{2\alpha x}+Ve^{\alpha x}$, with $-\infty<x<+\infty$ [6,7]. Therefore, the continuous energy scattering states are written as the infinite bounded sum of Eq. (3) with the expansion coefficients as the continuous dual Hahn polynomials $S_n^{\rho+\frac{1}{2}}\left(\varepsilon;\frac{\nu+1}{2},\frac{\nu+1}{2}\right)$. The requirement that the corresponding basis gives a tridiagonal matrix representation for the wave operator, $-\frac{1}{2}\frac{d^2}{dx^2}+\frac{\alpha^2}{8}e^{2\alpha x}+Ve^{\alpha x}-E$, results in (see section II.B.1 in Ref. [4])

$$\phi_n(x)=\sqrt{\tfrac{\Gamma(n+1)}{\Gamma(n+\nu+1)}}\,y^{\frac{\nu+1}{2}}e^{-y/2}L_n^\nu(y), \tag{18}$$

where $y(x)=e^{\alpha x}$. On the other hand, if $\rho<-\frac{1}{2}$ (i.e., $V<-\alpha^2/4$) then $\mu$ is negative and the problem has both continuous as well as discrete energy states and the corresponding wavefunction is given by Eq. (6). The discrete energy spectrum is obtained using formula (B8) as

$$E_k=-\tfrac{1}{2}\alpha^2\left(k+\tfrac{1}{2}+2V/\alpha^2\right)^2, \tag{19}$$

for $k=0,1,..,N$ and $N$ is the larger integer less than or equal to $-\left(2V/\alpha^2\right)-\frac{1}{2}$. This spectrum formula is identical to (14) above with $V_0=\alpha^2/8$ and $V_1=V$. Thus, we obtained here a full solution to the 1D Morse problem for both scattering and bound states whereas the solution obtained in the previous section was only for bound states.

### 3.2 The oscillator problem

Next, we make the parameter assignments: $\mu=\frac{1}{2}-\varepsilon$, $a=b=\frac{\nu+1}{2}$ and $z=\frac{\mathrm{i}}{2}\left(\ell+\frac{1}{2}\right)$. Thus, we obtain only bound states whose energy spectrum is given by formula (B8) as

$$E_k=\alpha^2\left(2k+\ell+\tfrac{3}{2}\right), \tag{20}$$

where we have chosen an inverse length parameter $\alpha$ and wrote $2E=\alpha^2\varepsilon$. This spectrum is identical to that of the 3D isotropic oscillator (11) above with oscillator frequency $\omega=\alpha^2$ and angular momentum quantum number $\ell$. The corresponding wavefunction is written as Eq. (5) in terms of the discrete dual Hahn polynomial. The requirement that the basis elements should result in a tridiagonal matrix representation for the wave operator, $-\frac{1}{2}\frac{d^2}{dr^2}+\frac{\ell(\ell+1)}{2r^2}+\frac{1}{2}\alpha^4r^2-E$, gives (see section II.B.2 in Ref. [4])

$$\phi_n(r)=\sqrt{\tfrac{\Gamma(n+1)}{\Gamma(n+\nu+1)}}\,(\alpha r)^{\nu+\frac{3}{2}}e^{-\alpha^2r^2/2}L_n^\nu(\alpha^2r^2). \tag{21}$$

### 3.3 The Coulomb problem

Finally, if we choose the parameters as: $\mu=\frac{1}{2}+\frac{\rho}{\sqrt{-\varepsilon}}$, $a=b=\frac{\nu+1}{2}$ and $z=\mathrm{i}\left(\ell+\frac{1}{2}\right)$ where $\varepsilon$ and $\rho$ are negative, then we obtain only bound states whose energy spectrum is given by formula (B8) as

$$E_k=-Z^2/2\left(k+\ell+1\right)^2, \tag{22}$$

where we have chosen an inverse length positive parameter $\lambda$ and wrote $2E=\lambda^2\varepsilon$, $\rho=Z/\lambda$. This is identical to the energy spectrum of the Coulomb problem given by Eq. (9) above with a negative electric charge $Z$ and angular momentum quantum number $\ell$. However, the solution here is only for bound states. The tridiagonal requirement on the basis functions results in the following realization (see section II.B. 3 in Ref. [4])

$$\phi_n(r)=\sqrt{\tfrac{\Gamma(n+1)}{\Gamma(n+\nu+1)}}\,(\lambda r)^{1+\frac{\nu}{2}}e^{-\lambda r/2}L_n^\nu(\lambda r), \tag{23}$$

with $\lambda^2=-8E_k$.

## 4. The Wilson polynomial class of problems

In this section, we consider the class of problems whose continuum scattering states are described by the wavefunction (3) where the expansion coefficients are the Wilson polynomials $W_n^\mu(z^2;\nu;a,b)$ shown in Appendix C by Eq. (C2). Throughout this section, we consider the special case where the two polynomial parameters $a$ and $b$ are equal and fixed by certain physical constraints (e.g., the number of bound states).

### 4.1 The hyperbolic Pöschl-Teller potential

We start by choosing the polynomial parameters as follows: $\nu+\mu=1+\sqrt{\frac{1}{4}+2u_1}$, $\nu-\mu=\sqrt{\frac{1}{4}-2u_0}$ and $z^2=\frac{1}{2}\varepsilon$, where all the parameters $\{\varepsilon,u_i\}$ are dimensionless and real with $u_0\le\frac{1}{8}$ and $u_1\ge-\frac{1}{8}$. If $\sqrt{\frac{1}{4}-2u_0}<1+\sqrt{\frac{1}{4}+2u_1}$, then $\mu$ is positive and the system consists only of continuous energy scattering states that are written in terms of $W_n^{\mu}\left(\frac{1}{2}\varepsilon;\nu;a,a\right)$. The phase shift associated with these scattering states is obtained using formula (C10) as

$$\begin{aligned}\delta(\kappa)&=\arg\Gamma\left(\mathrm{i}\,\kappa/\lambda\right)-\arg\Gamma\left(\tfrac{1}{2}+\tfrac{1}{2}\sqrt{\tfrac{1}{4}+2V_1/\lambda^2}-\tfrac{1}{2}\sqrt{\tfrac{1}{4}-2V_0/\lambda^2}+\tfrac{\mathrm{i}}{2}\tfrac{\kappa}{\lambda}\right)\\&-\arg\Gamma\left(\tfrac{1}{2}+\tfrac{1}{2}\sqrt{\tfrac{1}{4}+2V_1/\lambda^2}+\tfrac{1}{2}\sqrt{\tfrac{1}{4}-2V_0/\lambda^2}+\tfrac{\mathrm{i}}{2}\tfrac{\kappa}{\lambda}\right)-2\arg\Gamma\left(a+\tfrac{\mathrm{i}}{2}\tfrac{\kappa}{\lambda}\right)\end{aligned}\tag{24}$$

where we have introduced an inverse length scale parameter $\lambda$ and wrote $E=\lambda^2\varepsilon$, $V_i=\lambda^2u_i$, and defined the wave number $\kappa=\sqrt{2E}$. On the other hand, if $\sqrt{\frac{1}{4}-2u_0}>1+\sqrt{\frac{1}{4}+2u_1}$ then $\mu$ is negative and the system consist of both scattering as well as bound states and the corresponding wavefunction is written as shown by Eq. (6). The spectrum formula (C11) results in the following bound states energy eigenvalues

$$E_k=-\frac{\lambda^2}{2}\left(2k+1+\sqrt{\tfrac{1}{4}+2V_1/\lambda^2}-\sqrt{\tfrac{1}{4}-2V_0/\lambda^2}\right)^2,\tag{25}$$

where $k=0,1,..,N$ and $N$ is the largest integer less than or equal to $\frac{1}{2}\sqrt{\frac{1}{4}-2V_0/\lambda^2}-\frac{1}{2}\sqrt{\frac{1}{4}+2V_1/\lambda^2}-\frac{1}{2}$. These results are identical to those of the 1D hyperbolic Pöschl-Teller potential $V(x)=\dfrac{V_0}{\cosh^2(\lambda x)}+\dfrac{V_1}{\sinh^2(\lambda x)}$ with $x\ge 0$ [7,8]. The requirement that the basis $\{\phi_n(x)\}$ must produce a tridiagonal and symmetric representation of the Schrödinger wave operator with this potential gives (see section III.B.1 in Ref. [4])

$$\phi_n(x)=A_n(1-y)^{\alpha}(1+y)^{\beta}P_n^{(2a-1,\mu+\nu-1)}(y),\tag{26}$$

where $A_n=\sqrt{\frac{2n+\mu+\nu+2a-1}{2^{\mu+\nu+2a-1}}\frac{\Gamma(n+1)\Gamma(n+\mu+\nu+2a-1)}{\Gamma(n+2a)\Gamma(n+\mu+\nu)}}$, $y(x)=2\tanh^2(\lambda x)-1$, $\alpha=a$, $2\beta=\mu+\nu-\frac{1}{2}$ and $P_n^{(\sigma,\tau)}(y)$ is the Jacobi polynomial.

### 4.2 The trigonometric Scarf potential

For this problem, we choose $\nu+\mu=1+\sqrt{u_++u_-+\frac{1}{4}}$, $\nu-\mu=2\sqrt{\varepsilon-u_0}$ and $z^2=-\frac{1}{4}\left(u_+-u_-+\frac{1}{4}\right)$ with $\left(u_+\pm u_-\right)\ge-\frac{1}{4}$. Thus, $z$ is pure imaginary and the system consists only of bound states. The spectrum formula (C11) gives

$$\varepsilon_k=u_0+\left(k+\tfrac{1}{2}+\tfrac{1}{2}\sqrt{u_++u_-+\tfrac{1}{4}}+\tfrac{1}{2}\sqrt{u_+-u_-+\tfrac{1}{4}}\right)^2.\tag{27}$$

where $k=0,1,2,..,\infty$ since $\mu$ is always negative for any $k$. Thus, the bound states are written as in Eq. (5) with the discrete version of the Wilson polynomial, which is the Racah polynomials, $R_n^N(k;\alpha,\beta,\gamma)$, defined in Appendix C as expansion coefficients and with $N\to\infty$. If we introduce the inverse length scale parameter $\lambda$ and write $\varepsilon=2E/\lambda^2$ and $u_i=2V_i/\lambda^2$, then this becomes the energy spectrum of the 1D trigonometric Scarf potential $V(x)=V_0+\frac{V_+-V_-\sin(\lambda x)}{\cos^2(\lambda x)}$ where $|x|\le\pi/2\lambda$ [5]. The associated basis that produces a tridiagonal matrix representation for the Schrödinger wave operator with this potential is given by Eq. (26) with $y(x)=\sin(\lambda x)$, $\alpha=a+\frac{1}{4}$ and $2\beta=\mu+\nu-\frac{1}{2}$ (see section III.B.2 in Ref. [4]).

**4.3 The hyperbolic Eckart potential**

Now, we make the following choice of polynomial parameters: $\nu+\mu=1+\sqrt{4u_1+1}$, $\nu-\mu=2\sqrt{-\varepsilon-u_0}$ and $z^2=\varepsilon$. If the parameters are such that $\mu$ is positive then the system consists only of continuous energy scattering states that are written in terms of $W_n^\mu(\varepsilon;\nu;a,a)$. The corresponding phase shift is obtained from formula (C10) as

$$\begin{aligned}\delta(\varepsilon)&=\arg\Gamma\left(\mathrm{i}2\sqrt{\varepsilon}\right)-\arg\Gamma\left(\tfrac{1}{2}+\sqrt{u_1+\tfrac{1}{4}}-\mathrm{i}\sqrt{\varepsilon+u_0}+\mathrm{i}\sqrt{\varepsilon}\right)\\&-\arg\Gamma\left(\tfrac{1}{2}+\sqrt{u_1+\tfrac{1}{4}}+\mathrm{i}\sqrt{\varepsilon+u_0}+\mathrm{i}\sqrt{\varepsilon}\right)-2\arg\Gamma\left(a+\mathrm{i}\sqrt{\varepsilon}\right)\end{aligned}, \tag{28}$$

On the other hand, if $\mu$ is negative then the system consists of both scattering as well as bound states and the corresponding wavefunction is written as shown by Eq. (6). The spectrum formula (C1) results in the following energy spectrum

$$\varepsilon_k=-\frac{1}{4}\left[k+\frac{\mu+\nu+1}{2}-\frac{u_0}{k+\frac{\mu+\nu+1}{2}}\right]^2-u_0. \tag{29}$$

If we introduce the inverse length scale parameter $\lambda$ and write $\varepsilon=2E/\lambda^2$ and $u_i=2V_i/\lambda^2$, then these results become identical to those of the 1D hyperbolic Eckart potential [7,9]

$$V(x)=\frac{1}{e^{\lambda x}-1}\left[V_0+\frac{V_1}{1-e^{-\lambda x}}\right]=\frac{V_1/4}{\sinh^2(\lambda x/2)}+\frac{V_0/2}{\tanh(\lambda x/2)}-\frac{V_0}{2}, \tag{30}$$

with $x\ge 0$. The basis functions corresponding to this problem are those given by Eq. (26) with $y(x)=1-2e^{-\lambda x}$, $\alpha=a$ and $2\beta=\mu+\nu$ (see section III.B.3 in Ref. [4]).

**4.4 The hyperbolic Rosen-Morse potential**

The fourth and final problem corresponds to the following selection of polynomial parameters: $\nu+\mu=B-A+\frac{1}{2}$, $\nu-\mu=B+A+\frac{1}{2}$ and $z^2=\varepsilon$, where $A$ and $B$ are real dimensionless parameters. Thus, if $A$ is negative then the system is in the continuum energy state but if $A$ is positive then the system is a mix of finite discrete bound states and a continuous energy of scattering states. The corresponding phase shift and energy spectrum are obtained as follows

$$\delta(\varepsilon)=\arg\Gamma\left(\mathrm{i}2\sqrt{\varepsilon}\right)-\arg\Gamma\left(-A+\mathrm{i}\sqrt{\varepsilon}\right)-\arg\Gamma\left(B+\tfrac{1}{2}+\mathrm{i}\sqrt{\varepsilon}\right)-2\arg\Gamma\left(a+\mathrm{i}\sqrt{\varepsilon}\right), \tag{31}$$

$$\varepsilon_k=-\left(k-A\right)^2, \tag{32}$$

where $k=0,1,..,N$ and $N$ is the largest integer less than or equal to $-\mu=A$. If we introduce an inverse length parameter $\lambda$ and write $\varepsilon=2E/\lambda^2$, then these results correspond to the hyperbolic Rosen-Morse potential [7]

$$\frac{2}{\lambda^2}V(x)=\frac{(B^2+A^2+A)-B(2A+1)\cosh(\lambda x)}{\sinh^2(\lambda x)}, \tag{33}$$

where $x \geq 0$. The basis functions corresponding to this problem are those given by Eq. (26) with $y(x) = \cosh(\lambda x)$.

## 5. Conclusion and discussion

We have shown that by writing the wavefunction in terms of orthogonal polynomials as shown in one of the three appropriate forms given by equations (3), (5) or (6), then all physical information about the system is obtained from the properties of these polynomials. However, such polynomials are required to have the asymptotic behavior shown in Eq. (4). We found that the Coulomb, oscillator and Morse potentials are associated with the Meixner-Pollaczek and continuous dual Hahn polynomial classes. We also found that other well-known exactly solvable problems correspond to the Wilson polynomial class that includes it discrete version, the Racah polynomial. These latter problems include, but not limited to, the hyperbolic Pöschl-Teller, Eckart, Rosen-Morse and trigonometric Scarf potentials. We conclude by making an important remark and discussing two relevant issues:

- In the process of identifying the quantum mechanical system associated with a given polynomial, it might seem that we have made an arbitrary choice of polynomial parameters. However, those choices are, in fact, unique and were made carefully to correspond to the conventional class of exactly solvable potentials. Had we made an alternative choice of parameters, then the corresponding quantum system would have been different and might not belong to the well-known class of exactly solvable problems. Nonetheless, such alternative choices must respect any constraints on the parameters (e.g., reality and ranges). For example, if we choose the Meixner-Pollaczek polynomial parameters as: $\cosh\theta = \frac{\kappa - \mu\lambda}{\kappa + \mu\lambda}$ and $z = i\ln(\kappa/\lambda)$, where $\kappa^2 = 2E$ and $\mu < 0$ then we would have obtained the following bound states energy spectrum using formula (A8)
$$E_k = \tfrac{1}{2}\lambda^2 e^{-2(k+\mu)}, \tag{34}$$
where $k = 0,1,..,N$ and $N$ is the largest integer such that $e^{-(N+\mu)} > -\mu$. The polynomial that enters in the bound state wavefunction expansion (5) is the discrete version of the Meixner-Pollaczek polynomial with finite spectrum, which is the Krawtchouk polynomial. The energy spectrum (34) does not correspond to any of the known exactly solvable potentials making such a system very appealing and motivates us to search for the associated potential function. That is, doing the inverse quantum mechanical problem: finding the potential function starting from the energy spectrum. Such a problem is highly non-trivial. However, the tridiagonal representation requirement results in a sever restriction on the space of solutions of this problem making it tractable. A procedure to accomplish that and obtain the potential function analytically or numerically was developed and applied in Ref. [10].

- There is another class of four-parameter orthogonal polynomials, which was not treated in the mathematics or physics literature, but it corresponds to new solvable physical problems. The class consists of three polynomials, one with a continuous spectrum designated as $H_n^{(\mu,\nu)}(z;\alpha,\theta)$ and two of its discrete version with finite and infinite spectra. So far, it is defined by its three-term recursion relation that reads as follows [11,12]

$$\begin{aligned}(\cos\theta)H_n^{(\mu,\nu)}(z;\alpha,\theta) &= \left\{z\sin\theta\left[\left(n+\frac{\mu+\nu+1}{2}\right)^2+\alpha\right]+\frac{\nu^2-\mu^2}{(2n+\mu+\nu)(2n+\mu+\nu+2)}\right\}H_n^{(\mu,\nu)}(z;\alpha,\theta)\\ &+\frac{2(n+\mu)(n+\nu)}{(2n+\mu+\nu)(2n+\mu+\nu+1)}H_{n-1}^{(\mu,\nu)}(z;\alpha,\theta)+\frac{2(n+1)(n+\mu+\nu+1)}{(2n+\mu+\nu+1)(2n+\mu+\nu+2)}H_{n+1}^{(\mu,\nu)}(z;\alpha,\theta)\end{aligned} \tag{35}$$

where $0\le\theta\le\pi$. It is a polynomial of degree *n* in *z* and $\alpha$, which is obtained for all *n* staring with $H_0^{(\mu,\nu)}(z;\alpha,\theta)=1$ and $H_1^{(\mu,\nu)}(z;\alpha,\theta)$, which is computed from (35) by setting $n=0$ and $H_{-1}^{(\mu,\nu)}(z;\alpha,\theta)\equiv 0$. Setting $z\equiv 0$ turns (35) into the recursion relation of the Jacobi polynomial $P_n^{(\mu,\nu)}(\cos\theta)$. Physical requirements dictate that $\mu$ and $\nu$ are greater than $-1$ and $z\in\mathbb{R}$. In section III.A of Ref. [4], this polynomial class was used in solving several physical problems. These were associated with either new potential functions or generalizations of exactly solvable potentials. Table 1 is a partial list of potential functions associated with this class showing also the coordinate transformation $y(x)$ that enters in the basis elements (26) which supports a tridiagonal matrix representation for the corresponding wave operator. Note that the addition of the $V_1$ term in all of these potential functions prevents them from being exactly solvable in the standard formulation of quantum mechanics.

- The same formulation outlined above could be extended to the wavefunction of several variables as long as the corresponding problem is completely separable. For example, in three-dimensional configuration space with spherical coordinates, if we can write the total wavefunction as $\psi(r,\theta,\varphi)=\frac{1}{r}R(r)\Theta(\theta)\Phi(\varphi)$ then we can apply the same formulation and write Eq. (3) as follows

$$R_E^\mu(r)=\sqrt{\rho^\mu(\varepsilon)}\sum_n P_n^\mu(\varepsilon)\phi_n(r)\,, \tag{36a}$$

$$\Theta_{E_\theta}^\mu(\theta)=\sqrt{\tilde{\rho}^\mu(\varepsilon_\theta)}\sum_n \tilde{P}_n^\mu(\varepsilon_\theta)\tilde{\phi}_n(\theta)\,, \tag{36b}$$

$$\Phi_{E_\varphi}^\mu(\varphi)=\sqrt{\hat{\rho}^\mu(\varepsilon_\varphi)}\sum_n \hat{P}_n^\mu(\varepsilon_\varphi)\hat{\phi}_n(\varphi)\,, \tag{36c}$$

where $\left\{\phi_n(r),P_n^\mu(\varepsilon)\right\}$, $\left\{\tilde{\phi}_n(\theta),\tilde{P}_n^\mu(\varepsilon_\theta)\right\}$, and $\left\{\hat{\phi}_n(\varphi),\hat{P}_n^\mu(\varepsilon_\varphi)\right\}$ are the radial, angular, and azimuthal basis and associated polynomials, respectively. Each basis set must produce a tridiagonal matrix representation for the corresponding wave operator

$$\left[-\frac{1}{2}\frac{d^2}{dr^2}+\frac{E_\theta}{r^2}+V_r(r)-E\right]R_E^\mu(r)=0\,, \tag{37a}$$

$$\left[-\frac{1}{2}\left(1-x^2\right)\frac{d^2}{dx^2}+x\frac{d}{dx}+\frac{E_\varphi}{1-x^2}+V_\theta(\theta)-E_\theta\right]\Theta_{E_\theta}^\mu(\theta)=0\,, \tag{37b}$$

$$\left[-\frac{1}{2}\frac{d^2}{d\varphi^2}+V_\varphi(\varphi)-E_\varphi\right]\Phi_{E_\varphi}^\mu(\varphi)=0\,, \tag{37c}$$

where $x=\cos\theta$. The corresponding 3D separable potential function is $V(r,\theta,\varphi)=V_r(r)+\frac{1}{r^2}\left[V_\theta(\theta)+\frac{1}{1-x^2}V_\varphi(\varphi)\right]$. As illustration, one may consult Ref. [13] where we have applied this procedure to obtain an exact solution for a problem of this type and we wrote two alternative solutions to Eq. (37b); one in terms of $H_n^{(\mu,\nu)}(z;\alpha,\theta)$ and another in terms of the Wilson polynomial. In Refs. [14,15], the same was done in two dimensions for another problem.

## Acknowledgements

We are grateful to H. Hassanabadi, H. Bahlouli, S-H Dong, and Z. H. Yamani for the careful review and improvements on the original version of the manuscript. The support by the Saudi Center for Theoretical Physics (SCTP) is highly appreciated.

## Appendix A: The Two-Parameter Meixner-Pollaczek Polynomial Class

The orthonormal version of this polynomial is written as follows (see pages 37-38 of Ref. [16])

$$P_n^{\mu}(z,\theta)=\sqrt{\tfrac{(2\mu)_n}{n!}}\; e^{in\theta}\,{}_2F_1\left(\begin{smallmatrix}-n,\mu+iz\\2\mu\end{smallmatrix}\middle|1-e^{-2i\theta}\right), \tag{A1}$$

where $(a)_n=a(a+1)(a+2)...(a+n-1)=\frac{\Gamma(n+a)}{\Gamma(a)}$, $z$ is the whole real line, $\mu>0$ and $0<\theta<\pi$. This is a polynomial in $z$ which is orthonormal with respect to the measure $\rho^{\mu}(z,\theta)dz$. That is,

$$\int_{-\infty}^{+\infty}\rho^{\mu}(z,\theta)P_n^{\mu}(z,\theta)P_m^{\mu}(z,\theta)dz=\delta_{nm}\,, \tag{A2}$$

where the normalized weight function is

$$\rho^{\mu}(z,\theta)=\frac{1}{2\pi\Gamma(2\mu)}(2\sin\theta)^{2\mu}e^{(2\theta-\pi)z}\left|\Gamma(\mu+iz)\right|^2. \tag{A3}$$

These polynomials satisfy the following symmetric three-term recursion relation

$$\begin{aligned}\left(z\sin\theta\right)P_n^{\mu}(z,\theta)&=-\left[\left(n+\mu\right)\cos\theta\right]P_n^{\mu}(z,\theta)\\&+\frac{1}{2}\sqrt{n(n+2\mu-1)}P_{n-1}^{\mu}(z,\theta)+\frac{1}{2}\sqrt{(n+1)(n+2\mu)}P_{n+1}^{\mu}(z,\theta)\end{aligned} \tag{A4}$$

The asymptotics ($n\to\infty$) is obtained as follows (see, for example, the Appendix in Ref. [1])

$$P_n^{\mu}(z;\theta)\approx\frac{2n^{-1/2}e^{\left(\frac{1}{2}\pi-\theta\right)z}}{(2\sin\theta)^{\mu}\left|\Gamma(\mu+iz)\right|}\cos\left[n\theta+\arg\Gamma(\mu+iz)-\mu\frac{\pi}{2}-z\ln(2n\sin\theta)\right], \tag{A5}$$

which is in the required general form given by Eq. (4) because the extra $n$-dependent term $z\ln(2n\sin\theta)$ in the argument of the cosine could be ignored relative to $n\theta$ since $\ln n\approx o(n)$ as $n\to\infty$. Therefore, the scattering amplitude and phase shift are obtained as follows

$$A^{\mu}(\varepsilon)=2e^{\left(\frac{1}{2}\pi-\theta\right)z}\Big/(2\sin\theta)^{\mu}\left|\Gamma(\mu+iz)\right|, \tag{A6}$$

$$\delta^{\mu}(\varepsilon)=\arg\Gamma(\mu+iz)\,. \tag{A7}$$

The scattering amplitude (A6) shows that a discrete infinite spectrum occurs if $\mu+iz=-k$, where $k=0,1,2,...$ Thus, the spectrum formula associated with this polynomial is

$$z_k^2=-(k+\mu)^2, \tag{A8}$$

and bound states are written as in Eq. (5) where the discrete version of the Meixner-Pollaczek polynomial is obtained by the substitution $z = \mathrm{i}(k+\mu)$ and $\theta \to \mathrm{i}\theta$ in Eq. (A1). The latter substitution is needed to maintain reality of the recursion relation (A4). In fact, and as expected, the substitution $\theta \to \mathrm{i}\theta$ makes the asymptotics of (A1) vanish due to the decaying exponential $e^{-n\theta}$. Making these substitutions gives the discrete version as the following normalized Meixner polynomial (see pages 45-46 in Ref. [16])

$$M_n^{\mu}(k;\beta) = \sqrt{\tfrac{(2\mu)_n}{n!}}\,\beta^{n/2}\,{}_2F_1\!\left(\begin{smallmatrix}-n,-k\\2\mu\end{smallmatrix}\middle|1-\beta^{-1}\right), \tag{A9}$$

where $\beta = e^{-2\theta}$ with $\theta > 0$ making $0 < \beta < 1$. The substitution $z = \mathrm{i}(k+\mu)$ and $\theta \to \mathrm{i}\theta$ in the recursion relation (A4) together with $2\cosh\theta = \frac{1}{\sqrt{\beta}} + \sqrt{\beta}$ and $2\sinh\theta = \frac{1}{\sqrt{\beta}} - \sqrt{\beta}$ transform the recursion relation (A4) to

$$\begin{aligned}(\beta-1)k\,M_n^{\mu}(k;\beta) &= -\left[n(1+\beta)+2\mu\beta\right]M_n^{\mu}(k;\beta)\\ &+\sqrt{n(n+2\mu-1)\beta}\,M_{n-1}^{\mu}(k;\beta)+\sqrt{(n+1)(n+2\mu)\beta}\,M_{n+1}^{\mu}(k;\beta)\end{aligned} \tag{A10}$$

The associated normalized discrete weight function is $\omega_k^{\mu}(\beta) = (1-\beta)^{2\mu}\frac{\Gamma(k+2\mu)\,\beta^k}{\Gamma(2\mu)\Gamma(k+1)}$. That is, $\sum_{k=0}^{\infty}\omega_k^{\mu}(\beta)M_n^{\mu}(k;\beta)M_m^{\mu}(k;\beta) = \delta_{n,m}$. Due to the exchange symmetry of *n* and *k* in ${}_2F_1\!\left(\begin{smallmatrix}-n,-k\\2\mu\end{smallmatrix}\middle|1-\beta^{-1}\right)$, the Meixner polynomial is self-dual satisfying the dual orthogonality relation $\sum_{k=0}^{\infty}M_k^{\mu}(n;\beta)M_k^{\mu}(m;\beta) = \delta_{n,m}/\omega_n^{\mu}(\beta)$. Now, if we further take $2\mu = -N$, where *N* is a non-negative integer, then the indices *n* and *k* in (A9) cannot be larger than *N* otherwise the hypergeometric function blows up. Thus, the discrete Meixner polynomial with an infinite spectrum becomes the discrete Krawtchouk polynomial with a finite spectrum whose normalized version reads (see pages 46-47 in Ref. [16]):

$$K_n^{N}(k;\gamma) = \sqrt{\tfrac{N!}{n!(N-n)!}}\left(\tfrac{\gamma}{1-\gamma}\right)^{n/2}{}_2F_1\!\left(\begin{smallmatrix}-n,-k\\-N\end{smallmatrix}\middle|\gamma^{-1}\right), \tag{A11}$$

where we wrote $\gamma^{-1} = 1-\beta^{-1}$ with $0 < \gamma < 1$ and $n,k = 0,1,..,N$. In writing (A11), we have used $(-N)_n = \frac{\Gamma(n-N)}{\Gamma(-N)} = (-1)^n\frac{\Gamma(N+1)}{\Gamma(N-n+1)}$. These substitutions change the recursion relation (A10) to

$$\begin{aligned}k\,K_n^{N}(k;\gamma) &= \left[N\gamma + n(1-2\gamma)\right]K_n^{N}(k;\gamma)\\ &-\sqrt{n(N-n+1)\gamma(1-\gamma)}\,K_{n-1}^{N}(k;\gamma)-\sqrt{(n+1)(N-n)\gamma(1-\gamma)}\,K_{n+1}^{N}(k;\gamma)\end{aligned} \tag{A12}$$

The associated normalized discrete weight function is $\omega_k^{N}(\gamma) = (1-\gamma)^{N-k}\frac{\Gamma(N+1)\,\gamma^k}{\Gamma(N-k+1)\Gamma(k+1)}$, which is easily obtained from that of the Meixner polynomial by the substitution $2\mu = -N$ and $\beta = -\gamma/(1-\gamma)$. Then, $\sum_{k=0}^{N}\omega_k^{N}(\gamma)K_n^{N}(k;\gamma)K_m^{N}(k;\gamma) = \delta_{n,m}$. The Krawtchouk polynomial is also self-dual and satisfy the dual orthogonality relation $\sum_{k=0}^{N}K_k^{N}(n;\gamma)K_k^{N}(m;\gamma) = \delta_{n,m}/\omega_n^{N}(\gamma)$.

# Appendix B: The Three-Parameter Continuous Dual Hahn Polynomial Class

The orthonormal version of this polynomial is (see pages 29-31 of Ref. [16])

$$S_n^\mu(z^2;a,b)=\sqrt{\frac{(\mu+a)_n(\mu+b)_n}{n!(a+b)_n}}\,{}_3F_2\left(\left.{-n,\mu+\mathrm{i}z,\mu-\mathrm{i}z \atop \mu+a,\mu+b}\right|1\right), \tag{B1}$$

where ${}_3F_2\left(\left.{a,b,c \atop d,e}\right|z\right)=\sum_{n=0}^{\infty}\frac{(a)_n(b)_n(c)_n}{(d)_n(e)_n}\frac{z^n}{n!}$ is the generalized hypergeometric function, $z>0$ and $\mathrm{Re}(\mu,a,b)>0$ with non-real parameters occurring in conjugate pairs. This is a polynomial in $z^2$ which is orthonormal with respect to the measure $\rho^\mu(z;a,b)dz$ where the normalized weight function reads as follows

$$\rho^\mu(z;a,b)=\frac{1}{2\pi}\frac{\left|\Gamma(\mu+\mathrm{i}z)\Gamma(a+\mathrm{i}z)\Gamma(b+\mathrm{i}z)/\Gamma(2\mathrm{i}z)\right|^2}{\Gamma(\mu+a)\Gamma(\mu+b)\Gamma(a+b)}. \tag{B2}$$

That is, $\int_0^\infty S_n^\mu(z^2;a,b)S_m^\mu(z^2;a,b)\rho^\mu(z;a,b)dz=\delta_{nm}$. However, if the parameters are such that $\mu<0$ and $a+\mu$, $b+\mu$ are positive or a pair of complex conjugates with positive real parts, then the polynomial will have a continuum spectrum as well as a finite size discrete spectrum and the polynomial satisfies the following generalized orthogonality relation (Eq. 1.3.3 in [16])

$$\begin{gathered}\int_0^\infty \rho^\mu(z;a,b)\,S_n^\mu(z^2;a,b)\,S_m^\mu(z^2;a,b)\,dz-2\tfrac{\Gamma(a-\mu)\Gamma(b-\mu)}{\Gamma(a+b)\Gamma(1-2\mu)}\times\\ \sum_{k=0}^{N}(-1)^k\left(k+\mu\right)\tfrac{(\mu+a)_k(\mu+b)_k(2\mu)_k}{(\mu-a+1)_k(\mu-b+1)_k k!}\,S_n^\mu(k;a,b)\,S_m^\mu(k;a,b)=\delta_{n,m}\end{gathered} \tag{B3}$$

where $S_n^\mu(k;a,b)\equiv S_n^\mu\left(-(k+\mu)^2;a,b\right)$ and $N$ is the largest integer less than or equal to $-\mu$. It satisfies the following symmetric three-term recursion relation

$$\begin{aligned}z^2S_n^\mu(z^2;a,b)&=\left[(n+\mu+a)(n+\mu+b)+n(n+a+b-1)-\mu^2\right]S_n^\mu(z^2;a,b)\\&-\sqrt{n(n+a+b-1)(n+\mu+a-1)(n+\mu+b-1)}\,S_{n-1}^\mu(z^2;a,b)\\&-\sqrt{(n+1)(n+a+b)(n+\mu+a)(n+\mu+b)}\,S_{n+1}^\mu(z^2;a,b)\end{aligned} \tag{B4}$$

The asymptotics ($n\to\infty$) is (see, for example, the Appendix in Ref. [1])

$$\begin{gathered}S_n^\mu(z^2;a,b)\approx\frac{2}{\sqrt{n}}\frac{\sqrt{\Gamma(\mu+a)\Gamma(\mu+b)\Gamma(a+b)}\left|\Gamma(2\mathrm{i}z)\right|}{\left|\Gamma(\mu+\mathrm{i}z)\Gamma(a+\mathrm{i}z)\Gamma(b+\mathrm{i}z)\right|}\times\\ \cos\left\{z\ln n+\arg\left[\Gamma(2\mathrm{i}z)/\Gamma(\mu+\mathrm{i}z)\Gamma(a+\mathrm{i}z)\Gamma(b+\mathrm{i}z)\right]\right\}\end{gathered} \tag{B5}$$

Noting that $\ln n\approx o(n^\xi)$ for any $\xi>0$, then this result is also in the required general form given by Eq. (4). Therefore, the scattering amplitude and phase shift are obtained as follows

$$A^\mu(\varepsilon)=\frac{2\sqrt{\Gamma(\mu+a)\Gamma(\mu+b)\Gamma(a+b)}}{\left|\Gamma(\mu+\mathrm{i}z)\Gamma(a+\mathrm{i}z)\Gamma(b+\mathrm{i}z)/\Gamma(2\mathrm{i}z)\right|}, \tag{B6}$$

$$\delta^\mu(\varepsilon)=\arg\Gamma(2\mathrm{i}z)-\arg\Gamma(\mu+\mathrm{i}z)-\arg\Gamma(a+\mathrm{i}z)-\arg\Gamma(b+\mathrm{i}z). \tag{B7}$$

The scattering amplitude (B6) shows that a discrete finite spectrum occur if $\mu+\mathrm{i}z=-k$, where $k=0,1,2,..,N$ and $N$ is the largest integer less than or equal to $-\mu$. Thus, the spectrum formula associated with this polynomial is

$$z_k^2=-(k+\mu)^2. \tag{B8}$$

Substituting $z=\mathrm{i}(k+\mu)$ in Eq. (B1) and redefining the parameters as $2\mu=\alpha+\beta+1$, $\mu+a=\alpha+1$, $\mu+b=-N$, we obtain the discrete version of this polynomial as the normalized dual Hahn polynomial (see pages 34-36 in Ref. [16])

$$R_n^N\left(z_k^2;\alpha,\beta\right)=\sqrt{\frac{(\alpha+1)_n(N-n+1)_n}{n!(N+\beta-n+1)_n}}\,{}_3F_2\left(\left.{-n,-k,k+\alpha+\beta+1 \atop \alpha+1,-N}\right|1\right), \tag{B9}$$

where $z_k=k+\frac{\alpha+\beta+1}{2}$, $n,k=0,1,2,..,N$ and either $\alpha,\beta>-1$ or $\alpha,\beta<-N$. The same substitution in (B4) yields the following recursion relation

$$\begin{aligned}\left(k+\tfrac{\alpha+\beta+1}{2}\right)^2 R_n^N &=\left[(n+\alpha+1)(N-n)+n(N+\beta+1-n)+\tfrac{1}{4}(\alpha+\beta+1)^2\right]R_n^N\\ &+\sqrt{n(n+\alpha)(N-n+1)(N-n+\beta+1)}R_{n-1}^N+\sqrt{(n+1)(n+\alpha+1)(N-n)(N-n+\beta)}R_{n+1}^N\end{aligned} \tag{B10}$$

The associated normalized discrete weight function is

$$\omega_k^N(\alpha,\beta)=(\beta+1)_N\frac{(2k+\alpha+\beta+1)(\alpha+1)_k(N-k+1)_k}{(k+\alpha+\beta+1)_{N+1}(\beta+1)_k k!}. \tag{B11}$$

Therefore, the orthogonality reads: $\sum_{k=0}^N\omega_k^N(\alpha,\beta)R_n^N(k;\alpha,\beta)R_m^N(k;\alpha,\beta)=\delta_{n,m}$. It also satisfies the dual orthogonality

$$\sum_{k=0}^N R_k^N(n;\alpha,\beta)R_k^N(m;\alpha,\beta)=\delta_{n,m}\big/\omega_n^N(\alpha,\beta). \tag{B12}$$

# Appendix C: The Four-Parameter Wilson Polynomial Class

A version of the Wilson polynomial could be defined as follows (see pages 24-26 of Ref. [16])

$$\tilde{W}_n^\mu(z^2;\nu;a,b)=\frac{(\mu+a)_n(\mu+b)_n}{(a+b)_n n!}\,{}_4F_3\left(\left.{-n,n+\mu+\nu+a+b-1,\mu+iz,\mu-iz \atop \mu+\nu,\mu+a,\mu+b}\right|1\right), \tag{C1}$$

where $z>0$. However, its orthonormal version reads as follows

$$\begin{aligned}W_n^\mu(z^2;\nu;a,b)&=\sqrt{\left(\frac{2n+\mu+\nu+a+b-1}{n+\mu+\nu+a+b-1}\right)\frac{(\mu+\nu)_n(a+b)_n(\mu+\nu+a+b)_n\,n!}{(\mu+a)_n(\mu+b)_n(\nu+a)_n(\nu+b)_n}}\,\tilde{W}_n^\mu(z^2;\nu;a,b)\\ &=\sqrt{\left(\frac{2n+\mu+\nu+a+b-1}{n+\mu+\nu+a+b-1}\right)\frac{(\mu+a)_n(\mu+b)_n(\mu+\nu)_n(\mu+\nu+a+b)_n}{(\nu+a)_n(\nu+b)_n(a+b)_n n!}}\,{}_4F_3\left(\left.{-n,n+\mu+\nu+a+b-1,\mu+iz,\mu-iz \atop \mu+\nu,\mu+a,\mu+b}\right|1\right)\end{aligned} \tag{C2}$$

If $\mathrm{Re}(\mu,\nu,a,b)>0$ and non-real parameters occur in conjugate pairs, then the orthogonality relation becomes $\int_0^\infty W_n^\mu(z^2;\nu;a,b)W_m^\mu(z^2;\nu;a,b)\rho^\mu(z;\nu;a,b)dz=\delta_{nm}$, where the normalized weight function is

$$\rho^\mu(z;\nu;a,b)=\frac{1}{2\pi}\frac{\Gamma(\mu+\nu+a+b)\left|\Gamma(\mu+iz)\Gamma(\nu+iz)\Gamma(a+iz)\Gamma(b+iz)/\Gamma(2iz)\right|^2}{\Gamma(\mu+\nu)\Gamma(a+b)\Gamma(\mu+a)\Gamma(\mu+b)\Gamma(\nu+a)\Gamma(\nu+b)}. \tag{C3}$$

However, if the parameters are such that $\mu<0$ and $\mu+\nu$, $\mu+a$, $\mu+b$ are positive or a pair of complex conjugates with positive real parts, then the polynomial will have a continuum spectrum as well as a finite size discrete spectrum and the polynomial satisfies the following generalized orthogonality relation (Eq. 1.1.3 in [16])

$$\begin{aligned}&\int_0^\infty\rho^\mu(z;\nu;a,b)W_n^\mu(z^2;\nu;a,b)\,W_m^\mu(z^2;\nu;a,b)\,dz-2\frac{\Gamma(\mu+\nu+a+b)\Gamma(\nu-\mu)\Gamma(a-\mu)\Gamma(b-\mu)}{\Gamma(-2\mu+1)\Gamma(a+b)\Gamma(a+\nu)\Gamma(b+\nu)}\times\\ &\sum_{k=0}^N(k+\mu)\frac{(2\mu)_k(\mu+\nu)_k(\mu+a)_k(\mu+b)_k}{(\mu-\nu+1)_k(\mu-a+1)_k(\mu-b+1)_k k!}W_n^\mu(k;\nu;a,b)W_m^\mu(k;\nu;a,b)=\delta_{n,m}\end{aligned} \tag{C4}$$

where $W_n^\mu(k;\nu;a,b) \equiv W_n^\mu(-(k+\mu)^2;\nu;a,b)$ and $N$ is the largest integer less than or equal to $-\mu$. The associated symmetric three-term recursion relation is:

$$z^2 W_n^\mu = \left[\frac{(n+\mu+\nu)(n+\mu+a)(n+\mu+b)(n+\mu+\nu+a+b-1)}{(2n+\mu+\nu+a+b)(2n+\mu+\nu+a+b-1)} + \frac{n(n+\nu+a-1)(n+\nu+b-1)(n+a+b-1)}{(2n+\mu+\nu+a+b-1)(2n+\mu+\nu+a+b-2)} - \mu^2\right] W_n^\mu$$
$$-\frac{1}{2n+\mu+\nu+a+b-2}\sqrt{\frac{n(n+\mu+\nu-1)(n+a+b-1)(n+\mu+a-1)(n+\mu+b-1)(n+\nu+a-1)(n+\nu+b-1)(n+\mu+\nu+a+b-2)}{(2n+\mu+\nu+a+b-3)(2n+\mu+\nu+a+b-1)}}\, W_{n-1}^\mu \quad \text{(C5)}$$
$$-\frac{1}{2n+\mu+\nu+a+b}\sqrt{\frac{(n+1)(n+\mu+\nu)(n+a+b)(n+\mu+a)(n+\mu+b)(n+\nu+a)(n+\nu+b)(n+\mu+\nu+a+b-1)}{(2n+\mu+\nu+a+b-1)(2n+\mu+\nu+a+b+1)}}\, W_{n+1}^\mu$$

The asymptotics ($n \to \infty$) is (see, for example, Appendix B in Ref. [17])

$$W_n^\mu(z^2;\nu;a,b) \approx 2\sqrt{\frac{2}{n}B(\mu,\nu,a,b)}\,\left|\mathcal{A}(iz)\right|\cos\left[2z\ln n + \arg\mathcal{A}(iz)\right], \quad \text{(C6)}$$

where

$$B(\mu,\nu,a,b) = \Gamma(\mu+\nu)\Gamma(a+b)\Gamma(\mu+a)\Gamma(\mu+b)\Gamma(\nu+a)\Gamma(\nu+b)/\Gamma(\mu+\nu+a+b), \quad \text{(C7)}$$

and

$$\mathcal{A}(z) = \Gamma(2z)/\Gamma(\mu+z)\Gamma(\nu+z)\Gamma(a+z)\Gamma(b+z). \quad \text{(C8)}$$

Noting that $\ln n \approx o(n^\xi)$ for any $\xi > 0$, then this result is in the required general form given by Eq. (4). Therefore, the scattering amplitude and phase shift are obtained as follows

$$A^\mu(\varepsilon) = \frac{2\sqrt{2B(\mu,\nu,a,b)}}{\left|\Gamma(\mu+\mathrm{i}z)\Gamma(\nu+\mathrm{i}z)\Gamma(a+\mathrm{i}z)\Gamma(b+\mathrm{i}z)/\Gamma(2\mathrm{i}z)\right|}, \quad \text{(C9)}$$

$$\delta^\mu(\varepsilon) = \arg\Gamma(2\mathrm{i}z) - \arg\left[\Gamma(\mu+\mathrm{i}z)\Gamma(\nu+\mathrm{i}z)\Gamma(a+\mathrm{i}z)\Gamma(b+\mathrm{i}z)\right]. \quad \text{(C10)}$$

The scattering amplitude in this asymptotics shows that a discrete finite spectrum occur if $\mu + \mathrm{i}z = -k$, where $k = 0,1,2,..,N$ and $N$ is the largest integer less than or equal to $-\mu$. Thus, the spectrum formula associated with this polynomial is

$$z_k^2 = -(k+\mu)^2. \quad \text{(C11)}$$

Substituting $z = \mathrm{i}(k+\mu)$ in Eq. (C1) and redefining the parameters as $\mu = \frac{1}{2}(\gamma+\delta+1)$, $\nu = \beta + \frac{1}{2}(\delta-\gamma+1)$, $a = \alpha - \frac{1}{2}(\gamma+\delta-1)$ and $b = \frac{1}{2}(\gamma-\delta+1)$ then we obtain the following discrete version of the Wilson polynomial

$$\tilde{R}_n^N(k;\alpha,\beta,\gamma) = \frac{(\alpha+1)_n(\gamma+1)_n}{(\alpha+\beta+N+2)_n n!}\,{}_4F_3\left(\left.\begin{matrix}-n,-k,n+\alpha+\beta+1,k-\beta+\gamma-N\\ \alpha+1,\gamma+1,-N\end{matrix}\right|1\right), \quad \text{(C12)}$$

which is a renormalized Racah polynomial (see pages 26-29 of Ref. [16]) and the parameter $\delta$ is related to the integer $N$ as $\delta = -(N+\beta+1)$. The same substitution results in the following three-term recursion relation

$$\tfrac{1}{4}(N+\beta-\gamma-2m)^2\,\tilde{R}_n^N =$$
$$\left[\tfrac{1}{4}(N+\beta-\gamma)^2 - \frac{(n-N)(n+\alpha+1)(n+\gamma+1)(n+\alpha+\beta+1)}{(2n+\alpha+\beta+1)(2n+\alpha+\beta+2)} - \frac{n(n+\beta)(n+\alpha+\beta-\gamma)(n+N+\alpha+\beta+1)}{(2n+\alpha+\beta)(2n+\alpha+\beta+1)}\right]\tilde{R}_n^N \quad \text{(C13)}$$
$$+\frac{(n+\alpha)(n+\beta)(n+\gamma)(n+\alpha+\beta-\gamma)}{(2n+\alpha+\beta)(2n+\alpha+\beta+1)}\,\tilde{R}_{n-1}^N + \frac{(n+1)(n-N)(n+\alpha+\beta+1)(n+N+\alpha+\beta+2)}{(2n+\alpha+\beta+1)(2n+\alpha+\beta+2)}\,\tilde{R}_{n+1}^N$$

The discrete orthogonality reads as follows

$$\sum_{k=0}^{N}\frac{2k+\gamma-\beta-N}{k+\gamma-\beta-N}\,\frac{(-N)_k(\alpha+1)_k(\gamma+1)_k(\gamma-\beta-N+1)_k}{(-\beta-N)_k(\gamma-\beta+1)_k(\gamma-\alpha-\beta-N)_k\,k!}\,\bar{R}_n^N(k;\alpha,\beta,\gamma)\bar{R}_m^N(k;\alpha,\beta,\gamma)$$
$$=\frac{n+\alpha+\beta+1}{2n+\alpha+\beta+1}\,\frac{(-\alpha-\beta-N-1)_N(\gamma-\beta-N+1)_N}{(-\beta-N)_N(\gamma-\alpha-\beta-N)_N}\,\frac{(\beta+1)_n(\alpha+\beta-\gamma+1)_n(\alpha+\beta+N+2)_n\,n!}{(-N)_n(\alpha+1)_n(\gamma+1)_n(\alpha+\beta+2)_n}\,\delta_{n,m} \quad \text{(C14)}$$

which is formula (1.2.2) in [16] and where

$$\bar{R}_n^N(k;\alpha,\beta,\gamma) = {}_4F_3\left(\begin{matrix}-n,-k,n+\alpha+\beta+1,k-\beta+\gamma-N\\ \alpha+1,\gamma+1,-N\end{matrix}\middle|1\right). \tag{C15}$$

Therefore, the discrete normalized weight function is

$$\rho^N(k;\alpha,\beta,\gamma) = \frac{(-\beta-N)_N(\gamma-\alpha-\beta-N)_N}{(-\alpha-\beta-N-1)_N(\gamma-\beta-N+1)_N}\times \frac{2k+\gamma-\beta-N}{k+\gamma-\beta-N}\frac{(-N)_k(\alpha+1)_k(\gamma+1)_k(\gamma-\beta-N+1)_k}{(-\beta-N)_k(\gamma-\beta+1)_k(\gamma-\alpha-\beta-N)_k k!} \tag{C16}$$

and the orthonormal version of the discrete Racah polynomial is

$$R_n^N(k;\alpha,\beta,\gamma) = \sqrt{\frac{2n+\alpha+\beta+1}{n+\alpha+\beta+1}\frac{(-N)_n(\alpha+1)_n(\gamma+1)_n(\alpha+\beta+2)_n}{(\beta+1)_n(\alpha+\beta-\gamma+1)_n(\alpha+\beta+N+2)_n n!}}\; {}_4F_3\left(\begin{matrix}-n,-k,n+\alpha+\beta+1,k-\beta+\gamma-N\\ \alpha+1,\gamma+1,-N\end{matrix}\middle|1\right) \tag{C17}$$

Thus,

$$\sum_{k=0}^{N}\rho^N(k;\alpha,\beta,\gamma)R_n^N(k;\alpha,\beta,\gamma)R_m^N(k;\alpha,\beta,\gamma) = \delta_{n,m}. \tag{C18}$$

# References:

[1] A. D. Alhaidari and M. E. H. Ismail, *Quantum mechanics without potential function*, J. Math. Phys. **56** (2015) 072107

[2] K. M. Case, *Orthogonal polynomials from the viewpoint of scattering theory*, J. Math. Phys. **15** (1974) 2166

[3] J. S. Geronimo and K. M. Case, *Scattering theory and polynomials orthogonal on the real line*, Trans. Amer. Math. Soc. **258** (1980) 467494

[4] A. D. Alhaidari, *Solution of the nonrelativistic wave equation using the tridiagonal representation approach*, J. Math. Phys. **58** (2017) 072104; the polynomial $G_n^{(\mu,\nu)}(z;\sigma)$ defined in this paper is identical to the Wilson polynomial as $W_n^{-\frac{\nu+1}{2}-\sqrt{-\sigma}}\left(\frac{1}{2}z;\frac{\nu+1}{2}+\sqrt{-\sigma};\frac{\mu+1}{2},\frac{\mu+1}{2}\right)$

[5] See, for example, R. De, R. Dutt and U. Sukhatme, *Mapping of shape invariant potentials under point canonical transformations*, J. Phys. A **25** (1992) L843

[6] P. C. Ojha, SO(2,1) *Lie algebra, the Jacobi matrix and the scattering states of the Morse oscillator*, J. Phys. A **21** (1988) 875

[7] A. Khare and U. P. Sukhatme, *Scattering amplitudes for supersymmetric shape-invariant potentials by operator methods*, J. Phys. A **21** (1988) L501

[8] A. Frank and K. B. Wolf, *Lie algebras for systems with mixed spectra. I. The scattering Pöschl-Teller potential*, J. Math. Phys. **26** (1985) 973

[9] R. K. Yadav, A. Khare and B. P. Mandal, *The scattering amplitude for rationally extended shape invariant Eckart potentials*, Phys. Lett. A **379** (2015) 67

[10] A. D. Alhaidari, *Establishing correspondence between the reformulation of quantum mechanics without a potential function and the conventional formulation*, arXiv:1703.08659 [quant-ph]

[11] A. D. Alhaidari, *Orthogonal polynomials derived from the tridiagonal representation approach*, arXiv:1703.04039v2 [math-ph]

[12] A. D. Alhaidari, *Open problem in orthogonal polynomials*, arXiv:1709.06081v2 [math-ph]

[13] A. D. Alhaidari, *Scattering and bound states for a class of non-central potentials*, J. Phys. A **38** (2005) 3409
[14] A. D. Alhaidari, *Charged particle in the field of an electric quadrupole in two dimensions*, J. Phys. A **40** (2007) 14843
[15] F. M. Fernández, *Bound states of a charged particle in the field of an electric quadrupole in two dimensions*, J. Math. Chem. **52** (2014) 1576
[16] R. Koekoek and R. Swarttouw, *The Askey-scheme of hypergeometric orthogonal polynomials and its q-analogues*, Reports of the Faculty of Technical Mathematics and Informatics, Number 98-17 (Delft University of Technology, Delft, 1998)
[17] A. D. Alhaidari and T. J. Taiwo, *Wilson-Racah Quantum System*, J. Math. Phys. **58** (2017) 022101

# Table Caption:

**Table 1:** Partial list of exactly solvable potential functions in the polynomial class associated with $H_n^{(\mu,\nu)}(z;\alpha,\theta)$. The coordinate transformation $y(x)$ enters in the basis (26) that supports a tridiagonal matrix representation for the corresponding wave operator. The presence of the $V_1$ term in all of these potentials inhibits exact solvability of the Schrödinger wave equation in the standard formulation of quantum mechanics.

**Table 1**

| $V(x)$ | $x$ | $y(x)$ |
|---|---|---|
| $V_0 + \dfrac{V_+ - V_- \sin(\pi x/L)}{\cos^2(\pi x/L)} + V_1 \sin\left(\pi x/L\right)$ | $-\tfrac{1}{2}L \le x \le +\tfrac{1}{2}L$ | $\sin\left(\pi x/L\right)$ |
| $\dfrac{1}{1-(x/L)^2}\left\{V_0 + \dfrac{V_+}{(x/L)^2} + \dfrac{V_-}{1-(x/L)^2} + V_1\left[2(x/L)^2 - 1\right]\right\}$ | $0 \le x \le L$ | $2\left(x/L\right)^2 - 1$ |
| $\dfrac{1}{e^{\lambda x}-1}\left[V_0 + V_- e^{\lambda x} + \dfrac{V_+}{1-e^{-\lambda x}} + V_1\left(1-2e^{-\lambda x}\right)\right]$ | $x \ge 0$ | $1-2e^{-\lambda x}$ |
| $V_- + \dfrac{V_+}{\sinh^2(\lambda x)} + \dfrac{V_0 + V_1[2\tanh^2(\lambda x) - 1]}{\cosh^2(\lambda x)}$ | $x \ge 0$ | $2\tanh^2(\lambda x) - 1$ |
| $V_+ - V_- \tanh(\lambda x) + \dfrac{V_0 + V_1 \tanh(\lambda x)}{\cosh^2(\lambda x)}$ | $-\infty < x < +\infty$ | $\tanh(\lambda x)$ |
| $V_0 + \dfrac{V_+}{\sin^2(\pi x/L)} + \dfrac{V_-}{\cos^2(\pi x/L)} - V_1 \cos(2\pi x/L)$ | $0 \le x \le \tfrac{1}{2}L$ | $2\sin^2(\pi x/L) - 1$ |
| $V_0 + \dfrac{V_+ - V_- \cosh(\lambda x)}{\sinh^2(\lambda x)} + V_1 \cosh\left(\lambda x\right)$ | $x \ge 0$ | $\cosh\left(\lambda x\right)$ |